\documentclass[aps,pre,twocolumn,amsmath,amssymb,superscriptaddress,showpacs,longbibliography]{revtex4-1}

\usepackage{graphicx}
\usepackage{dcolumn}
\usepackage{bm}
\usepackage{mathrsfs}

\usepackage{natbib}
\usepackage[text={7in,9.5in},centering]{geometry}

\usepackage{hyperref}
\usepackage{url}

\usepackage{epsfig}
\usepackage{amsmath}
\usepackage{amssymb}
\usepackage{textcomp}

\usepackage{color}
\usepackage{float}

\usepackage{stackengine}
\usepackage{verbatim}

\begin{document}

\title{Localization of nonbacktracking centrality on dense subgraphs of sparse networks}

\author{G. Tim\'ar}
 \email{gtimar@ua.pt}
 \affiliation{Departamento de F\'\i sica da Universidade de Aveiro \& I3N, Campus Universit\'ario de Santiago, 3810-193 Aveiro, Portugal}

\author{S. N. Dorogovtsev}
 \affiliation{Departamento de F\'\i sica da Universidade de Aveiro \& I3N, Campus Universit\'ario de Santiago, 3810-193 Aveiro, Portugal}

\author{J. F. F. Mendes}
 \affiliation{Departamento de F\'\i sica da Universidade de Aveiro \& I3N, Campus Universit\'ario de Santiago, 3810-193 Aveiro, Portugal}

\date{\today}

\begin{abstract}
The nonbacktracking matrix, and the related nonbacktracking centrality (NBC) play a crucial role in models of percolation-type processes on networks, such as non-recurrent epidemics. Here we study the localization of NBC in infinite sparse networks that contain an arbitrary finite subgraph. Assuming the local tree-likeness of the enclosing network, and that branches emanating from the finite subgraph do not intersect at finite distances, we show that the largest eigenvalue of the nonbacktracking matrix of the composite network is equal to the highest of the two largest eigenvalues: that of the finite subgraph and of the enclosing network. In the localized state, when the largest eigenvalue of the subgraph is the highest of the two, we derive explicit expressions for the NBCs of nodes in the subgraph and other nodes in the network. In this state, nonbacktracking centrality is concentrated on the subgraph and its immediate neighbourhood in the enclosing network. We obtain simple, exact formulas in the case where the enclosing network is uncorrelated. We find that the mean NBC decays exponentially around the finite subgraph, at a rate which is independent of the structure of the enclosing network, contrary to what was found for the localization of the principal eigenvector of the adjacency matrix. Numerical simulations confirm that our results provide good approximations even in moderately sized, loopy, real-world networks.
\end{abstract}

\maketitle

\section{Introduction}
\label{sec1}

Recent decades have witnessed a surge of scientific activity focused on understanding the behaviour of dynamical models on complex network substrates. Much of this research effort has been oriented towards models of epidemic spreading on networks of social acquaintances, such as the paradigmatic susceptible-infected-susceptible (SIS) endemic model, and the susceptible-infected-removed (SIR) model of non-recurrent epidemics \cite{pastor2015epidemic, newman2002spread}. Mean-field theories have been developed for both classes of models, making use of two characteristic matrices: the adjacency matrix and the non-backtracking (or Hashimoto) matrix \cite{hashimoto2014automorphic} of a given network, respectively. In the quenched mean field approximation of the SIS model on a static network, the epidemic threshold is given as the inverse of the largest eigenvalue (LEV) of the adjacency matrix \cite{wang2003epidemic, castellano2010thresholds, ferreira2012epidemic}. Close to the threshold the probability of a given node being infected is proportional to the corresponding component of the principal eigenvector (PEV). This has the consequence that close to the epidemic threshold, disease may become localized on the largest hubs and their immediate neighbourhoods \cite{goltsev2012localization}. These results are partly an artefact of the quenched mean field approximation, which is not exact even in infinite locally tree-like networks due to neglecting dynamical correlations between the infection states of neighbouring nodes. Nevertheless the predicted disease localization does also occur in real-life systems \cite{dorogovtsev2022nature, kulldorff1995spatial, tanser2009localized, ferreira2016metastable}, and is a crucial phenomenon to understand in epidemic surveillence and control \cite{herrera2016disease, valdez2012intermittent, zhang2014cluster}.

A similar approach for non-recurrent epidemics, analogous to the quenched mean field in the case of the SIS model, is the message-passing approximation where it is assumed that the contribution of a node $j$ to the behaviour of a neighbouring node $i$ is completely determined by the contribution of the neighbours of $j$, excluding $i$. This assumption holds exactly for infinite locally tree-like networks in the case of percolation \cite{karrer2014percolation} and, e.g., the SIR epidemic model \cite{karrer2010message, shrestha2015message}. The relevant matrix in message-passing theories is the non-backtracking (NB) matrix $\mathbf{H}$, which is a $2L \times 2L$ nonsymmetric matrix ($L$ being the number of links in the network) whose elements are indexed by directed links $i \leftarrow j$, instead of nodes. It is defined as $H_{i \leftarrow j, k \leftarrow l} = \delta_{j,k} (1 - \delta_{i,l})$, where $\delta$ is the Kronecker symbol.

Similarly to the quenched mean field in the case of the SIS model, message-passing predicts a percolation or SIR epidemic threshold that is the inverse of the LEV of the NB matrix \cite{shrestha2015message}. Also, the probability, close to the threshold, of node $i$ belonging to the giant connected component (or of being infected in an SIR epidemic) is proportional to the nonbacktracking centrality (NBC) of node $i$ (see Ref. \cite{kuhn2017heterogeneous}), defined as $\textrm{NBC}_i \equiv \sum_{j \in \partial_i} v_{i \leftarrow j}$, where $\partial_i$ is the set of neighbours of node $i$ and $v_{i \leftarrow j}$ is the component of the PEV of the NB matrix that corresponds to the directed link $i \leftarrow j$. In other words, $\textrm{NBC}_i$ is the sum of incoming PEV components to node $i$. We will assume the NBC values to be normalized according to the condition $\sum_i^N \textrm{NBC}_i = 1$, in a network consisting of $N$ nodes. The NBC emerges in various applications related to non-recurrent dynamical models, e.g., in designing optimal percolation and immunization strategies \cite{morone2015influence, morone2016collective, torres2020node} or identifying influential spreaders \cite{radicchi2016leveraging, min2018identifying}.
Compared to the classical eigenvector centrality (the PEV components of the adjacency matrix), NBC suffers to a much lesser degree from localization, as discussed in Ref. \cite{martin2014localization}. In particular, isolated hubs cannot be centers of localization, that is, the NBC is able to avoid the artificial ``self-inflating'' phenomenon associated with high-degree nodes in the case of the adjacency matrix PEV. This circumstance has also made the NB matrix a useful tool in spectral community detection methods \cite{krzakala2013spectral, bordenave2015non}.

Although the PEV of the NB matrix cannot be localized on isolated hubs, localization may still occur on dense subgraphs, e.g., cliques \cite{martin2014localization}, if the subgraph in question has a LEV higher than that of the surrounding network. A recent study of a large set of real-world networks showed that the presence of such dense subgraphs is quite common, in which case the LEV of the network is dominated by the LEV of the subgraph and nodes outside the subgraph have NBC close to zero \cite{pastor2020localization}.
In particular, the authors of Ref. \cite{pastor2020localization} identified two specific subgraphs in real-world networks that were often found to be centers of localization of the NBC: the maximum $k$-core and a bipartite structure called \textit{overlapping hubs}.
These subgraphs may correspond to localized percolation clusters, and localized disease spreading in the SIR model. In such cases typically a double transition is found, where the first transition is determined by the structure of the dense subgraph and the second by the entire network \cite{pastor2020localization}. The message-passing method is only able to identify the first transition, and treats it as a strict phase transition, which is an artifact of this approach \cite{timar2017nonbacktracking}.
As an extreme example, consider a small clique with high LEV embedded in a larger network with smaller LEV. Message-passing will predict a single phase transition at the inverse of the LEV of the small clique. This is not a true phase transition, in the sense that susceptibility here is still small, of the order of $1$. It is rather an indication that localized cooperative phenomena may already occur, namely the clique and its neighbourhood may experience a small epidemic outbreak, in the case of the SIR model.
To be able to interpret message-passing results correctly it is vital to have a solid understanding of such localization phenomena.

In the present work we explore the localization of the NBC in a more general setting than previously considered, deriving some exact results and useful approximations. The paper is organized as follows.
In Sec. \ref{sec2} we describe our network construction which allows for the derivation of exact results: an arbitrary finite subgraph embedded in an arbitrary infinite locally tree-like network, where branches emanating from the finite subgraph do not intersect at finite distances.
In Sec. \ref{sec3}, using the concept of nonbacktracking expansion, we derive an exact expression for the NBCs of nodes in the finite subgraph (\textit{child network}).
We show that the LEV of the composite network is exactly given by the maximum of the two LEV values: that of the child network and that of the embedding infinite locally tree-like network (\textit{mother network}). Thus localization occurs when the composite LEV is equal to the LEV of the child.
We obtain compact approximate formulas for the NBC values in the case where the embedding mother network is uncorrelated. Performing simulations we check that our predictions also work well with a real-world mother network with degree correlations and short loops.
In Sec. \ref{sec4} we show that in the localized state the mean NBC decays exponentially with distance from the child network. Interestingly the rate of decay depends only on the LEV of the child network, and is completely independent of the mother network. We show that in finite mother networks the exponential form has a cutoff distance depending on network size and the difference between the LEVs of child and mother network. We give a summary and conclusions in Sec. \ref{sec5}.

\section{Arbitrary graph inserted in an infinite locally tree-like network}
\label{sec2}

We aim to quantify the localization of the NBC on arbitrarily structured small, dense subgraphs embedded in large sparse networks. Our setup is somewhat idealized, enabling us to derive exact results and simple approximate expressions. Nonetheless, as we show in Sections \ref{sec3} and \ref{sec4} in various examples, the strict assumptions can be greatly relaxed and the theory still provides good predictions.

Throughout our derivations we will make extensive use of the concept of nonbacktracking expansion of graphs, introduced in \cite{timar2017nonbacktracking}, which corresponds to computational trees in computer science \cite{weiss2001optimality}. For a given graph $\mathcal{G}$, the nonbacktracking expansion provides an infinite tree constructed as follows. Starting from an arbitrary node $i$ in $\mathcal{G}$ as the root of the tree, perform all possible non-backtracking walks from this node.
Each nonbacktracking walk corresponds to a branch of the infinite tree.
Thus the nonbacktracking expansion ``unfolds'' the original graph into an infinite tree, generating an infinite number of replicas for each node in $\mathcal{G}$.
It was shown in \cite{timar2017nonbacktracking} that for an arbitrary graph, the LEV of the NB matrix corresponds to the asymptotic branching of the nonbacktracking expansion.
(The branching of an infinite tree-like network is defined as the limit $\lim_{\ell\to\infty} L_{\ell+1}/L_\ell$, where $L_\ell$ is the number of links connecting the nodes at distance $\ell$ from an arbitrary node to the nodes at distance $\ell+1$, see Ref. \cite{lyons1989ising}.)
Furthermore, the NBC of node $i$ in graph $\mathcal{G}$ is equal to the relative frequency of replicas of node $i$ on the surface of the tree at infinity, irrespective of the starting node (root). This useful property was applied recently in Ref. \cite{timar2021approximating} to approximate NBC values of nodes in correlated networks.

\begin{figure}[H]
\centering
\includegraphics[width=\columnwidth,angle=0.]{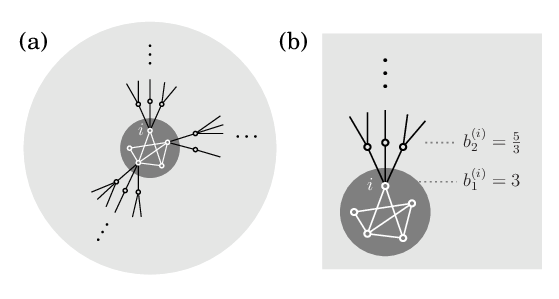}
\caption{(a) A finite child network (white nodes and links on dark grey shaded area) inserted into an infinite locally tree-like mother network (black nodes and links on light grey shaded area). The branches emanating from nodes of the child network do not intersect within any finite distance. In this example the number of nodes selected for merging with mother network nodes is $n_c=3$. (b) First two elements in the sequence of branching numbers on the branch connected to node $i$ of the child network.}
\label{fig:insertion}
\end{figure}

Let us consider a system in which an arbitrary finite child network of $n$ nodes is inserted into an infinite locally tree-like network in the following way. Each node $i$, of $n_c \leq n$ selected nodes in the child network, is merged with a node of degree $b_1^{(i)}$ in the mother network.
By merging we mean that the resulting node (of the composite network) retains the links of both the child node and the mother node. As an example, see Fig. \ref{fig:insertion}(a) where child node $i$ of degree 2 is merged with a mother node of degree 3, resulting in a node of degree 5 in the composite network.
Let the sequence of branchings on the mother network branches attached to node $i$ be $b_2^{(i)}, b_3^{(i)}, \ldots, b_m^{(i)}, \ldots$ [see Fig. \ref{fig:insertion}(b)], where $b_{m \to \infty}^{(i)} \to b$. Here $b$ is the branching of the mother network, i.e., the LEV of its NB matrix. For all nodes $j$ of the child network that were not chosen for merging with mother network nodes, $b_1^{(j)}, b_2^{(j)}, b_3^{(j)}, \ldots = 0$. Importantly, the branches attached to different nodes in the child network do not intersect within any finite distance. Fig. \ref{fig:insertion} demonstrates this construction. The non-intersection of branches emanating from child nodes is a strong assumption, which certainly does not hold in real-world networks that tend to have many short loops. As we will see, the theory also provides good approximations in these more realistic scenarios.

\noindent
Now let us denote the LEV of the NB matrix of a child network (to be inserted into a mother network) by $z$. This is the asymptotic branching of the nonbacktracking expansion of this subgraph. This means that asymptotically the number of vertices on the surface of the expansion at distance $m$ is

\begin{align}
H_m \cong C z^m,
\label{eq2.10}
\end{align}

\noindent
where $C$ is a constant. Let the NBC of node $i$ in the child network be $x_i$, $i=1,2,\ldots,n$, normalized according to $\sum_{i=1}^n x_i = 1$.

\section{Nonbacktracking centralities in the child network}
\label{sec3}

\subsection{Results for general networks}
\label{sec30}

Let us perform the nonbacktracking expansion of the composite network, starting from one of the child nodes (see Ref. \cite{timar2017nonbacktracking} for details of this procedure). The number of replica nodes of the child network on the surface of the nonbacktracking expansion at large distances $m$ will still be $Cz^m$, but there will be additional vertices from the mother network. We want to find the fraction $\rho$ of replica nodes of the child network on the surface of the expansion as $m \to \infty$. This is equal to the sum of NBCs of child nodes. (We denote the NBC of child node $i$ by $\textrm{NBC}_i$. Note that this value---the NBC measured in the composite network---differs from $x_i$.) Let $Q_m$ be the total number of nodes on the surface of the expansion at distance $m$. In the limit of infinite $m$,

\begin{align}
\rho = \lim_{m \to \infty} \frac{Cz^m}{Q_m}.
\label{eq3.10}
\end{align}

\noindent
For large $m$ we have the following recursion relation,

{
\medmuskip=0mu
\thinmuskip=0mu
\thickmuskip=0mu
\begin{align}
Q_m = & Cz^m + Cz^{m-1} \sum_{i=1}^n x_i b_1^{(i)} + Cz^{m-2} \sum_{i=1}^n x_i b_1^{(i)} b_2^{(i)} + \nonumber \\
&+ Cz^{m-3} \sum_{i=1}^n x_i b_1^{(i)} b_2^{(i)} b_3^{(i)} + \ldots .
\label{eq3.20}
\end{align}
}

\noindent
Let us explain this equation. The first term on the right-hand side is the number of nonbacktracking walks of length $m$ that stay inside the child network throughout the $m$ steps. The second term is the number of nonbacktracking walks of length $m$ that spend $m-1$ steps inside the child network and exit it only in the last step. In general, the $i^{\textrm{th}}$ term gives the number of nonbacktracking walks of length $m$ that spend $m-i+1$ steps inside the child network and exit it in the next step, never returning due to the non-intersecting tree nature of all branches emanating from the child network.
Eqs. (\ref{eq3.10}) and (\ref{eq3.20}) lead to the following relation,

{
\medmuskip=0mu
\thinmuskip=0mu
\thickmuskip=0mu
\begin{align}
\frac{1}{\rho} &= 1 + \sum_{i=1}^n x_i \frac{b_1^{(i)}}{z} + \sum_{i=1}^n x_i \frac{b_1^{(i)} b_2^{(i)} }{z^2} + \sum_{i=1}^n x_i \frac{b_1^{(i)} b_2^{(i)} b_3^{(i)} }{z^3} + \ldots.
\label{eq3.30}
\end{align}
}

\noindent
Hence we have,

\begin{align}
\rho = \left[  \sum_{i=1}^n x_i \left(  1 + \frac{b_1^{(i)}}{z} + \frac{ b_1^{(i)} b_2^{(i)} }{z^2} + \frac{ b_1^{(i)} b_2^{(i)} b_3^{(i)} }{z^3} + \ldots \right)   \right]^{-1}.
\label{eq3.40}
\end{align}

\noindent
We immediately see from Eq. (\ref{eq3.40}) that $\rho = 0$ if $b \geq z$, which implies that, in this case, the LEV of the composite network is $\lambda_{\textrm{cmp}} = b$. On the other hand, in the localized state, when $z>b$, the sum in parentheses converges, resulting in $\rho > 0$. In this case, as we show in Sec. \ref{sec4}, almost all nodes outside the child network have negligible NBC: only the nodes of the child network and nodes close to it have NBC noticably different from zero. Since the branches emanating from the child network do not intersect, no finite neighbourhood of the child network contains additional loops (compared to those already present in the child network), and so the subgraph comprising the nodes with non-negligible NBCs has the same LEV as the child network, and hence, $\lambda_{\textrm{cmp}} = z$. Therefore, we have in general that

\begin{align}
\lambda_{ \textrm{cmp} } = \text{max}( z, b ).
\label{eq3.45}
\end{align}

\begin{figure}[H]
\centering
\includegraphics[width=\columnwidth,angle=0.]{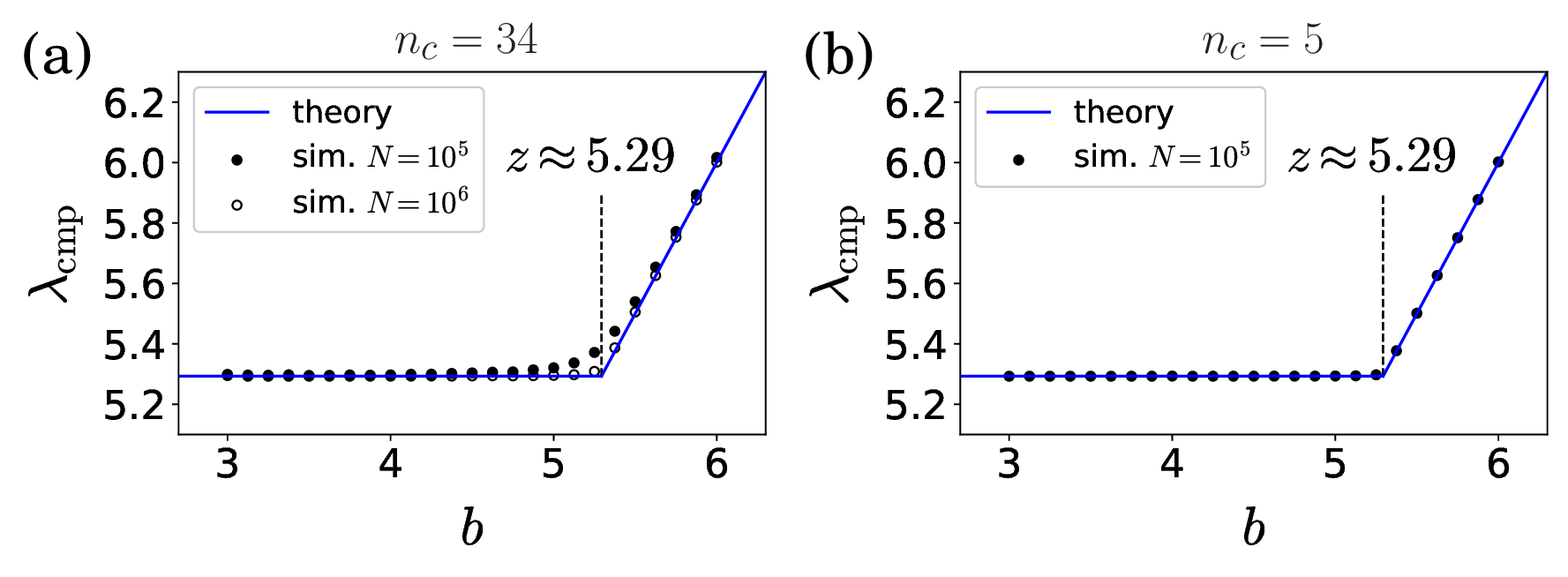}
\caption{LEV of the composite network consisting of the Karate club network ($n=34, z\approx5.29$) inserted into an ER network of varying mean degree $\langle k \rangle \approx b$. Solid black dots correspond to simulation results for a mother network size of $N=10^5$, and empty black dots correspond to a mother network size of $N=10^6$. The solid blue line corresponds to the result for $N \to \infty$, Eq. (\ref{eq3.45}). (a) The Karate club network was inserted into the ER mother network by merging all $n_c=n=34$ nodes with random nodes of the mother network. (b) Only $n_c=5$ random nodes of the Karate club network were merged with random nodes of the mother network. Simulation results were averaged over $100$ realizations for each point in both panels.}
\label{fig:LEV_figures}
\end{figure}

\noindent
Making use of the Collatz-Wielandt formula, a corollary of the Perron-Frobenius theorem, one can easily show that the LEV of any network is greater than or equal to the LEV of an arbitrary subgraph of the network (see, e.g., Ref. \cite{martin2014localization}). According to Eq. (\ref{eq3.45}) this inequality is actually a strict equality in the case of an arbitrary child network in an infinite locally tree-like mother network.
Equation (\ref{eq3.45}) is confirmed in Fig. \ref{fig:LEV_figures}, where, as an example, the small \textit{Karate club} network \cite{network_source} is inserted into a large Erd\H os-R\'enyi (ER) network of varying mean degree. (An ER mother network was chosen for convenience, as it is locally tree-like, and its LEV is closely approximated by its mean degree, for large sizes.) The insertion is made in a way that a number $n_c$ of nodes of the child network are merged with $n_c$ randomly chosen nodes in the mother network. The LEV of the composite network was found to follow Eq. (\ref{eq3.45}) very closely for small $n_c$, where the condition of non-intersecting branches is a decent approximation. For larger $n_c$ there is less ``space'' in the mother network for the branches to be independent, making the theory less accurate, and resulting in a smoother crossover of the composite LEV. This is a finite size effect, and would disappear for $N \to \infty$ (see Fig. \ref{fig:LEV_figures}a). A rough criterion for the validity of the results derived for $N \to \infty$ is given in Sec. \ref{sec33}.

\noindent
We can write Eq. (\ref{eq3.40}) in simpler form.
Since $b_{m \to \infty}^{(i)} \to b$, we have, asymptotically,

\begin{align}
\prod_{k=1}^m b_k^{(i)} \cong c_i b^m,
\label{eq3.50}
\end{align}

\noindent
where $c_i$ is a constant,

\begin{align}
c_i \equiv \lim_{m \to \infty} \frac{1}{b^m} \prod_{k=1}^m b_k^{(i)}.
\label{eq3.60}
\end{align}

\noindent
The function $\rho^{-1}$ contains a singularity at $z=b$, hidden in Eqs. (\ref{eq3.30}) and (\ref{eq3.40}). Extracting this singularity we obtain

{
\medmuskip=0mu
\thinmuskip=0mu
\thickmuskip=0mu
\begin{align}
\rho &= \Bigg[  \sum_{i=1}^n x_i \Bigg( c_i + 1 - c_i + \frac{ c_i b + b_1^{(i)} - c_i b }{z} + \nonumber \\
&+ \frac{ c_i b^2 + b_1^{(i)} b_2^{(i)} - c_i b^2  }{z^2} + \frac{ c_i b^3 + b_1^{(i)} b_2^{(i)} b_3^{(i)} - c_i b^3 }{z^3} + \ldots \Bigg)   \Bigg]^{-1} \nonumber \\
&= \Bigg[  \sum_{i=1}^n x_i \Bigg( \frac{c_i}{1-b/z} + 1 - c_i + \frac{ b_1^{(i)} - c_i b }{z} + \nonumber \\
&+ \frac{ b_1^{(i)} b_2^{(i)} - c_i b^2  }{z^2} + \frac{ b_1^{(i)} b_2^{(i)} b_3^{(i)} - c_i b^3 }{z^3} + \ldots \Bigg)   \Bigg]^{-1},
\label{eq3.70}
\end{align}
}

\noindent
or, in more compact form,


\begin{align}
\rho = &\Bigg\{ \frac{ \sum_{i=1}^n x_i c_i }{ z-b } z  + \sum_{i=1}^n x_i \Bigg[ 1 - c_i + \nonumber \\
&+ \sum_{k=1}^{\infty} \frac{1}{z^k} \Bigg(  \prod_{j=1}^k b_j^{(i)} - c_i b^k  \Bigg)   \Bigg]  \Bigg\}^{-1}.
\label{eq3.80}
\end{align}

\noindent
Note that the only assumption for the mother network was local tree-likeness. The mother network may possess any correlations, as long as tree-likeness is fulfilled; it may be random or deterministic.

The NBC of a given node $i$ of the child network is simply 

\begin{align}
\textrm{NBC}_i = x_i \rho.
\label{eq3.85}
\end{align}

\noindent
This is easily confirmed by noticing that the fraction of replicas of node $i$, among the replicas of all child nodes, on the surface of the nonbacktracking expansion at infinity is $x_i$ regardless of the structure of the surrounding mother network.

Throughout the rest of the paper, unless explicitly stated otherwise, we will always consider the localized state, i.e., when $z>b$.

\subsection{Approximation for finite distances}
\label{sec31}

Equation (\ref{eq3.80}) is an exact result, given the somewhat strict assumptions of an infinite locally tree-like mother network, and non-intersecting branches emanating from child nodes. To turn this result into a useful expression in real networks, we consider distances only up to a finite $m$ value. One can measure the branching numbers $b_1^{(i)}, \, b_2^{(i)}, \, \ldots$ up to $b_m^{(i)}$, for all child nodes $i$, and assume that the branching numbers are equal to $b$ thereafter. The constants $c_i$ are in this case defined as

\begin{align}
c_i \equiv \frac{1}{b^m} \prod_{k=1}^m b_k^{(i)},
\label{eq3.90}
\end{align}

\noindent
and the formula for $\rho$ changes to

\begin{align}
\rho \approx &\Bigg\{ \frac{ \sum_{i=1}^n x_i c_i }{ z-b } z  + \sum_{i=1}^n x_i \Bigg[ 1 - c_i + \nonumber \\
&+ \sum_{k=1}^{m-1} \frac{1}{z^k} \Bigg(  \prod_{j=1}^k b_j^{(i)} - c_i b^k  \Bigg)   \Bigg]   \Bigg\}^{-1}.
\label{eq3.100}
\end{align}

\noindent
We can derive a first order approximation by considering just one step away from the child network, i.e., $m=1$,

\begin{align}
\rho \approx \frac{z-b}{ \sum_{i=1}^n x_i (z-b+k_i) },
\label{eq3.110}
\end{align}

\noindent
where $k_i = b_1^{(i)}$ is the degree of the mother network node with which the child node $i$ was merged.

\subsection{Random mother network}
\label{sec32}

Informative expressions can be found for average quantities when the child network is inserted into a random uncorrelated network with an arbitrary degree distribution. In this case we can obtain $\langle \rho^{-1} \rangle$ exactly, averaged over the members of the configuration model ensemble, using Eq. (\ref{eq3.80}). In this expression only $c_i$ and $b_j^{(i)}$ are random variables, and they appear independently, so we can perform the average:

\begin{align}
\langle \rho^{-1} \rangle = & \frac{ \sum_{i=1}^n x_i \langle c_i \rangle }{ z-b } z  + \sum_{i=1}^n x_i \Bigg[ 1 - \langle c_i \rangle + \nonumber \\
&+ \sum_{k=1}^{\infty} \frac{1}{z^k} \Bigg(  \left\langle \prod_{j=1}^k b_j^{(i)} \right\rangle  - \langle c_i \rangle b^k  \Bigg)   \Bigg] .
\label{eq32.10}
\end{align}

\noindent
Here, for an uncorrelated network, the branching $b = \langle k^2 \rangle / \langle k \rangle - 1$, where $\langle k \rangle$ and $\langle k^2 \rangle$ are the first and second moments of the mother network's degree distribution. (We assume a finite second moment.) Note that in uncorrelated networks the sequence $b_1^{(i)}, b_2^{(i)}, b_3^{(i)}, \ldots$ almost surely converges to $b$ for any $i$.
Also, $b_1^{(i)}, b_2^{(i)}, b_3^{(i)}, \ldots, b_k^{(i)}$ are independent random variables with mean values $\langle k \rangle, b, b, \ldots, b$. Therefore, $\langle \prod_{j=1}^k b_j^{(i)} \rangle = \langle k \rangle b^{k-1}$, for any $k \geq 1$, and hence, using Eq. (\ref{eq3.60}), we have that $\langle c_i \rangle = \langle k \rangle / b$. Using these observations Eq. (\ref{eq32.10}) can be written as

\begin{align}
\langle \rho^{-1} \rangle = \frac{ \langle k \rangle + z - b }{ z - b }.
\label{eq32.20}
\end{align}

\noindent
Note the similarity between Eq. (\ref{eq32.20}) and the first order approximation for general mother networks, Eq. (\ref{eq3.110}).
Assuming that the variance of $\rho$ is small, we can estimate

\begin{align}
\langle \rho \rangle \approx \frac{ z - b}{ \langle k \rangle + z-b }.
\label{eq32.30}
\end{align}

\noindent
By Jensen's inequality we know that

\begin{align}
\langle \rho \rangle \geq \langle \rho^{-1} \rangle^{-1},
\label{eq32.35}
\end{align}

\noindent
so the expression in Eq. (\ref{eq32.30}) is a lower bound on the exact value. Equality holds only if the variance of $\rho$ is zero. Simulations show that this approximation is fairly accurate.

\begin{figure}[H]
\centering
\includegraphics[width=\columnwidth,angle=0.]{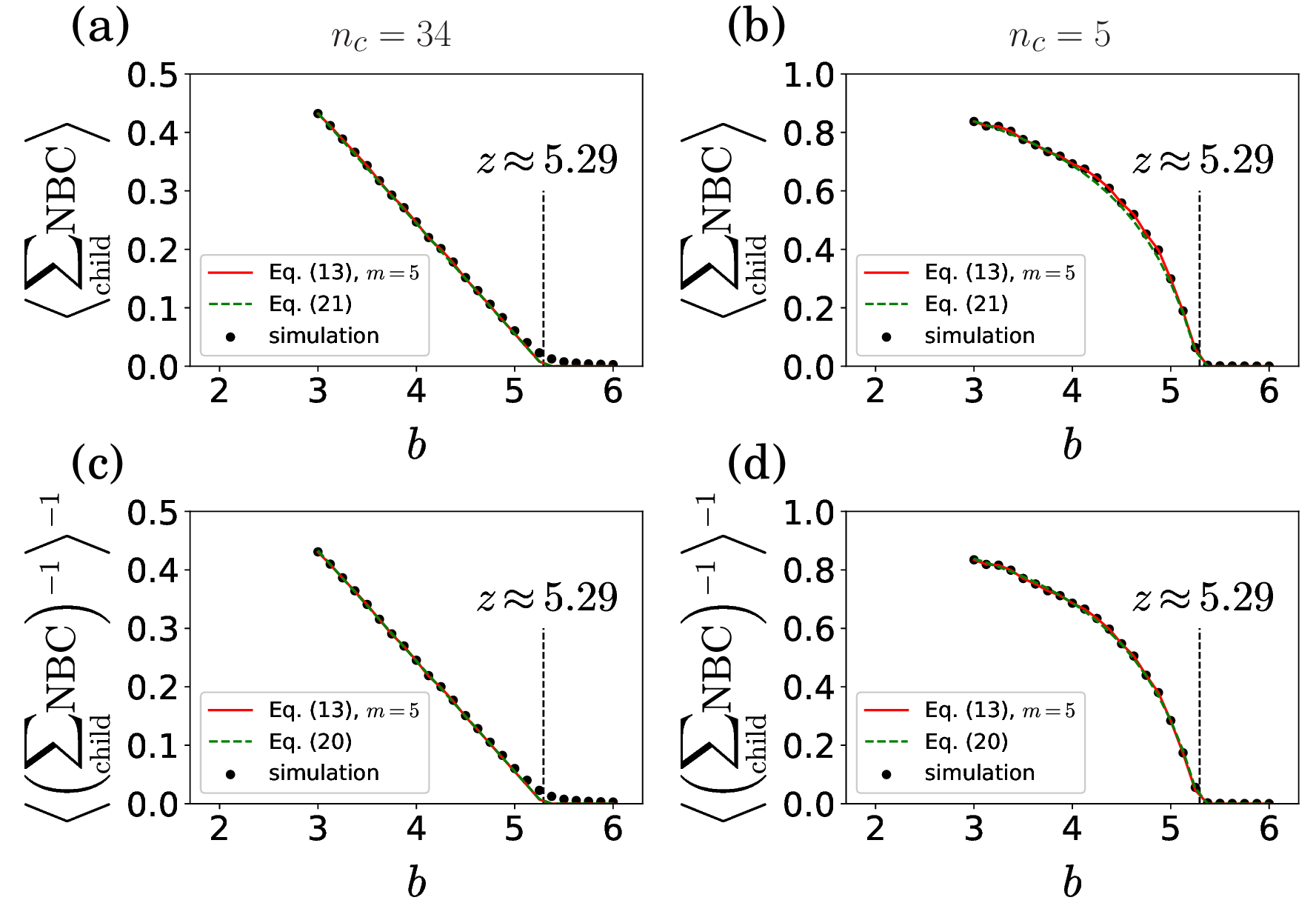}
\caption{The Karate club network ($n=34, z\approx5.29$) inserted into an ER network of $N=10^5$ with varying mean degree $\langle k \rangle \approx b$.  (a),(b) Average sum of NBCs of child nodes, and (c),(d) inverse of average inverse sum of NBCs of child nodes, as functions of the LEV of the mother network. The solid red line corresponds to the result Eq. (\ref{eq3.100}), evaluated up to a distance $m=5$. The dashed green line corresponds to the results for random mother networks, Eqs. (\ref{eq32.40}) and (\ref{eq32.50}). (a),(c) The Karate club network was inserted into the ER mother network by merging all $n_c=n=34$ nodes with random nodes of the mother network. (b),(d) Only $n_c=5$ random nodes of the Karate club network were merged with random nodes of the mother network. Simulation results were averaged over $100$ realizations for each point in all panels.}
\label{fig:rho_figures}
\end{figure}

For an ER mother network, $\langle k \rangle = b$, so we have the simple relationship

\begin{align}
\langle \rho^{-1} \rangle = \frac{z}{z-b},
\label{eq32.37}
\end{align}

\noindent
which is readily observed in Fig. \ref{fig:rho_figures}.
Equations (\ref{eq32.20}) and (\ref{eq32.37}) are completely general exact results, for an arbitrary child network, given that all nodes of the child network are merged with random mother network nodes.
If only a random fraction $f$ of nodes in the child network is merged with random mother network nodes, then

\begin{align}
\langle \rho^{-1} \rangle = \frac{ f \langle k \rangle + z-b }{z - b},
\label{eq32.40}
\end{align}

\noindent
and hence, again assuming a small variance of $\rho$, we have

\begin{align}
\langle \rho \rangle \approx \frac{ z - b}{ f \langle k \rangle + z-b }.
\label{eq32.50}
\end{align}

\noindent
As an example, Fig. \ref{fig:rho_figures} shows results for the sum of NBCs of nodes in the Karate club network inserted into an ER network of size $N=10^5$. The exact formula, Eq. (\ref{eq32.40}), fits the simulation results perfectly when the number of connection nodes $n_c$ is small. For larger $n_c$ deviations from the theory are observed close to the localization transition due to the intersection of branches already at short distances. For decreasing mean degree of the mother network the formula works progressively better, as expected. Eq. (\ref{eq32.50}) also works remarkably well, despite being strictly only a lower bound. This supports our assumption of a small variance for the distribution of $\rho$.

\subsection{Range of validity of the theory}
\label{sec33}

We can give a rough criterion for the validity of the results derived in the limit $N \to \infty$, considering the normalization condition for the NBCs, $\sum_{i=1}^N \textrm{NBC}_i = 1$. As we show in Sec. \ref{sec4}, in the localized state the NBC of nodes in the mother network decays exponentially with distance from the child network. This means that nodes outside the child network, on average, have NBC much smaller than $\rho / n$, if the network is large. Together with the normalization condition, for large networks, this means that the network size $N$ must satisfy

\begin{align}
\frac{\rho}{n} N \gg 1.
\label{eq33.10}
\end{align}

\noindent
Using Eqs. (\ref{eq33.10}) and (\ref{eq32.40}), and assuming $z \approx b$, we arrive at a rough criterion for the validity of the theory,

\begin{align}
\frac{N}{n_c} \gg \frac{ \langle k \rangle }{z-b}.
\label{eq33.20}
\end{align}

Eq. (\ref{eq33.20}) shows that, as $z$ approaches $b$, $N$ must increase approximately as $\sim (z-b)^{-1}$ for the theory to remain valid. Conversely, for a fixed network size, the theory becomes less accurate close to the point $z=b$, which is confirmed by Figs. \ref{fig:LEV_figures} and \ref{fig:rho_figures}.

\subsection{Real-world mother network}
\label{sec34}

\begin{figure}[H]
\centering
\includegraphics[width=\columnwidth,angle=0.]{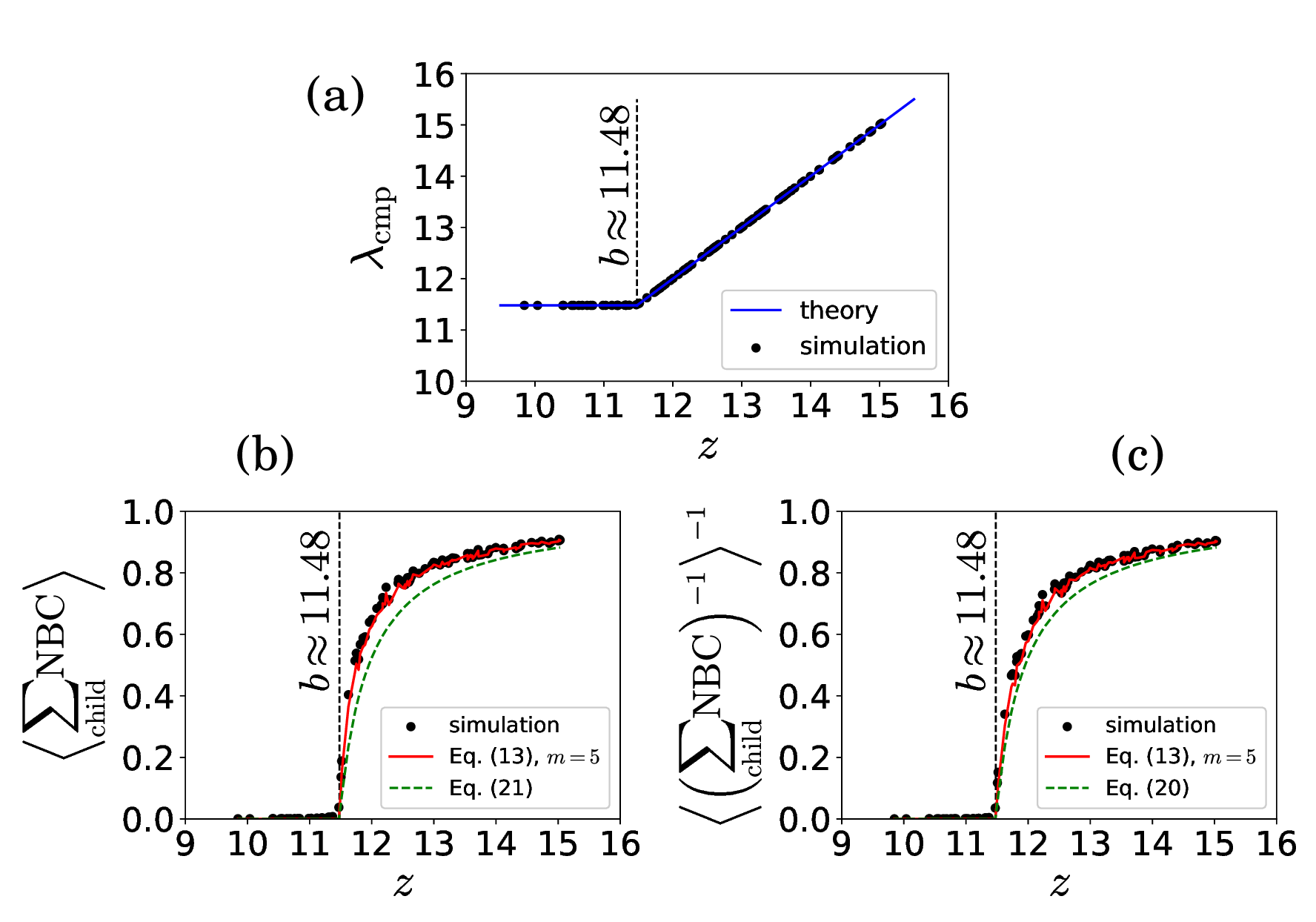}
\caption{An ER network of $n=50$ and varying mean degree, inserted into the Gnutella p2p network ($N=62561, b \approx 11.48$) at $n_c=5$ random nodes. (a) LEV of the composite network as a function of the LEV of the child network. (b) Average sum of NBCs of child nodes. (c) Inverse of average inverse sum of NBCs of child nodes. Simulation results were averaged over $100$ realizations for each point in all panels.}
\label{fig:ER_in_real}
\end{figure}

Many real-world networks, particularly social networks, tend to have a multitude of short loops, violating our assumption of non-intersecting branches emanating from the child network. Simulation results indicate that Eq. (\ref{eq3.100}) still works well, and even the results derived for uncorrelated random networks, Eqs. (\ref{eq32.40}) and (\ref{eq32.50}) provide decent approximations (see Fig. \ref{fig:ER_in_real}). As an example, a small ($n=50$) ER network is inserted into the \textit{Gnutella p2p} network \cite{network_source} by merging $n_c=5$ random nodes of the child network with random mother network nodes, and the sum of NBCs of the child network is compared with our predictions. (An ER network was, again, chosen only for convenience.)

\section{Exponential decay of nonbacktracking centrality around the child network}
\label{sec4}

\subsection{Results for general networks}
\label{sec40}

Let us find how nonbacktracking centrality is distributed in the composite network.
Returning to the nonbacktracking expansion of the composite network, let $\zeta_m^{(i)}$ be the fraction of replicas---on the surface of the expansion---of nodes at distance $m \geq 1$ from child node $i$ (on the branch attached to node $i$). This is the sum of the NBCs of these nodes. For $z>b$ Eq. (\ref{eq3.20}) and (\ref{eq3.30}) immediately lead to the following expression:

\begin{align}
\zeta_m^{(i)} = \frac{\rho}{ z^m } x_i \prod_{k=1}^m b_k^{(i)} \cong \rho c_i x_i \left( \frac{b}{z} \right)^m.
\label{eq4.10}
\end{align}

\noindent
According to Eq. (\ref{eq3.40}),

\begin{align}
\rho + \sum_{i=1}^n \sum_{m=1}^{\infty} \zeta_m^{(i)} = 1,
\label{eq4.20}
\end{align}

\noindent
as is natural. Furthermore, since the number of nodes in the branches attached to node $i$ at distance $m$ from this node grows as $\sim c_i b^m$, the NBC for an ``average node'' at distance $m$ decays as

\begin{align}
\frac{ \zeta_m^{(i)} }{ c_i b^m } \cong \rho x_i z^{-m},
\label{eq4.30}
\end{align}

\noindent
which is an exponential decay with rate $z$. Notably the branching of the mother network doesn't play a role here.
Using Eq. (\ref{eq4.10}) the sum of NBCs of all nodes at a distance $m \geq 1$ from the child network can be written as

\begin{align}
S_m = \frac{\rho}{ z^m } \sum_{i=1}^n x_i \prod_{k=1}^m b_k^{(i)} \cong \rho \left( \frac{b}{z} \right)^m \sum_{i=1}^n c_i x_i,
\label{eq4.40}
\end{align}

\noindent
while the number of nodes at distance $m \geq 1$ from the child network,

\begin{align}
N_m = \sum_{i=1}^n \prod_{k=1}^m b_k^{(i)} \cong b^m \sum_{i=1}^n c_i.
\label{eq4.50}
\end{align}

\noindent
So the mean NBC of nodes at a distance $m \geq 1$ from the child network is

\begin{align}
\langle \textrm{NBC} \rangle_m = \frac{S_m}{N_m} \cong \rho z^{-m} \frac{ \sum_{i=1}^n c_i x_i }{ \sum_{i=1}^n c_i }.
\label{eq4.60}
\end{align}

\noindent
If the child network is a regular graph, then $x_i = 1/n$, and

\begin{align}
\langle \textrm{NBC} \rangle_m = \frac{S_m}{N_m} \cong \frac{ \rho z^{-m} }{n}.
\label{eq4.70}
\end{align}



These results are similar to the case of the adjacency matrix PEV in a localized state, studied in Ref. \cite{goltsev2012localization}. The authors also found an exponential decay of eigenvector centrality as a function of the distance from the localization centre: the node of highest degree. The authors of Ref. \cite{goltsev2012localization} considered a less general situation, a hub of degree $q$ inserted into a Bethe lattice of branching $B$, and found that the adjacency matrix PEV components decay with distance $m$ as $f_m \propto a^{-m}$, with $a = (q-B)^{1/2}$, and the critical value for localization $q_{\textrm{loc}} = B^2 + B$. In other words the rate of decay depends on the branching of the surrounding network, whereas, interestingly, in the case of the NBC the decay rate depends only on the LEV of the child network. On the other hand, as the degree of the hub $q$ approaches $q_{\textrm{loc}}$ from above, $a$ tends to $B$. Thus, at the localization transition both the eigenvector centrality and NBC decay at a rate given by the branching of the mother network.

Localization in a given vector $\vec{h} = (h_1, h_2, ..., h_N)$ is often studied using the inverse participation ratio, $Y_4 (\vec{h}) = (\sum_i^N h_i^4) / (\sum_i^N h_i^2)^2$, observing how this quantity scales with system size $N$. For the case of NBCs in our study, when $z < b$, we have $Y_4 (\textrm{NBC}) \sim N^{-1}$, indicating a delocalized state. For $z>b$, considering that the decay of NBC around the child network is exponential, we obtain $Y_4 (\textrm{NBC}) \sim \textrm{const}$, indicating that localization occurs on a finite subgraph, as expected.

\subsection{Random mother network}
\label{sec41}

If the mother network is a random uncorrelated network, we can calculate some averages approximately. In this case let us define an average NBC at distance $m$ as

\begin{align}
\overline{ \textrm{NBC} }_m \equiv \frac{ \langle S_m \rangle }{ \langle N_m \rangle },
\label{eq41.10}
\end{align}

\noindent
where $\langle S_m \rangle$ and $\langle N_m \rangle$ are the averages over the network ensemble. We have

\begin{align}
\langle S_m \rangle \approx f \langle \rho \rangle \langle k \rangle \frac{ b^{m-1} }{ z^m }, \label{eq41.20} \\
\langle N_m \rangle = fn \langle k \rangle b^{m-1},
\label{eq41.30}
\end{align}

\noindent
where $f$ is the fraction of child nodes that are merged with mother network nodes. (In Eq. (\ref{eq41.20}) we made the approximation $\langle \rho \prod_{k=1}^m b_k^{(i)} \rangle \approx \langle \rho \rangle \langle \prod_{k=1}^m b_k^{(i)} \rangle$, again assuming a small variance for $\rho$.) We obtain

\begin{align}
\overline{ \textrm{NBC} }_m = \frac{ \langle S_m \rangle }{ \langle N_m \rangle } \approx \frac{ \langle \rho \rangle z^{-m} }{ n }.
\label{eq41.40}
\end{align}

\begin{figure}[H]
\centering
\includegraphics[width=\columnwidth,angle=0.]{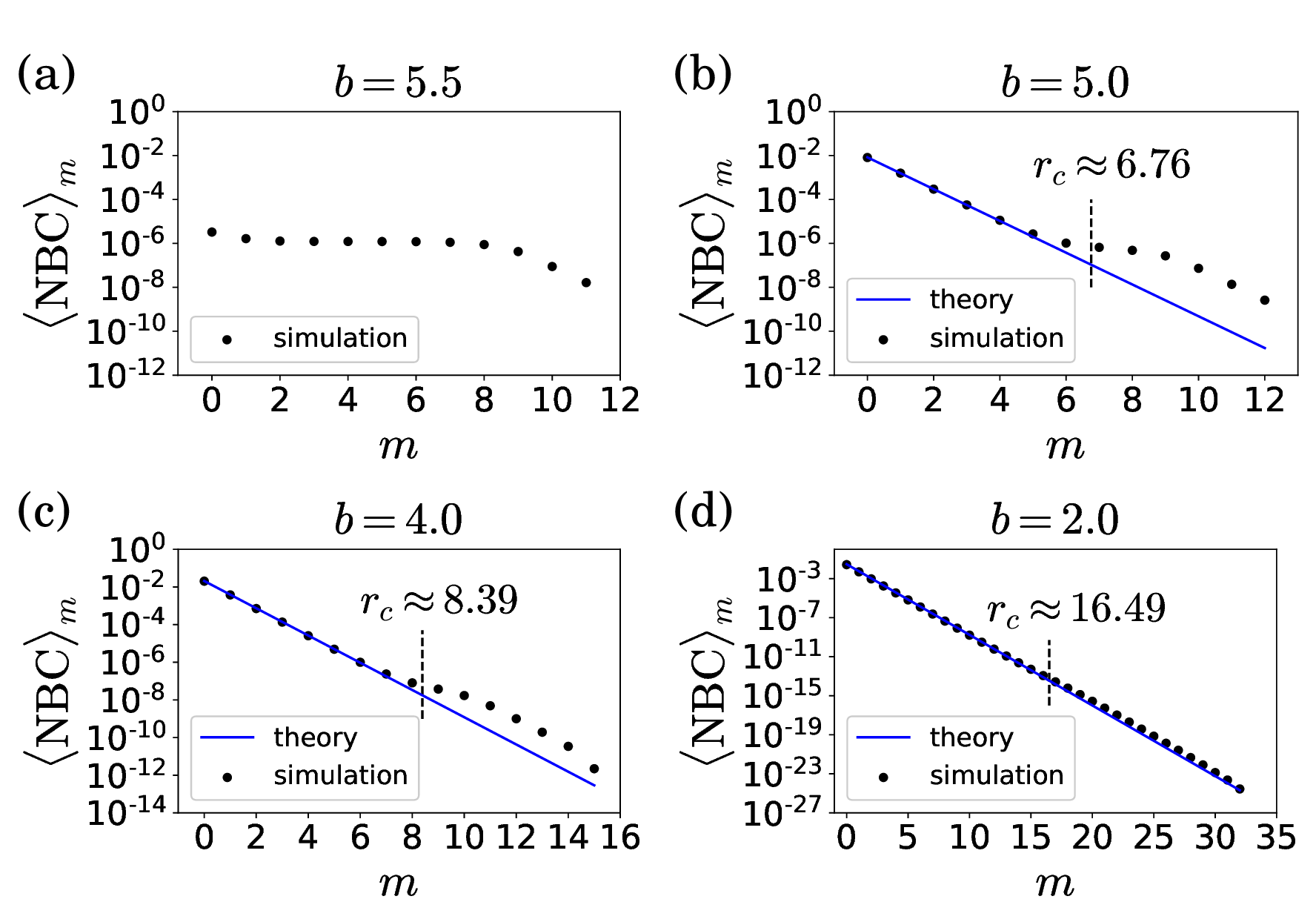}
\caption{Exponential decay of NBC around the child network: the Karate club network ($n=34, z\approx5.29$) inserted into an ER network of $N=10^6$ with varying mean degree $\langle k \rangle \approx b$. The insertion was made at $n_c=5$ random points. Black dots show mean NBC as a function of distance from the child network, for different LEV values of the ER mother network. Solid blue lines correspond to the result, Eq. (\ref{eq41.40}). The range of exponential decay, $r_c$, given by Eq. (\ref{eq42.50}) is marked with dashed black lines. In panel (a) localization of the NBC is absent, and the mean NBC is almost uniform. Simulation results were averaged over $10$ realizations for each point in all panels.}
\label{fig:decay_figures}
\end{figure}

\begin{figure}[H]
\centering
\includegraphics[width=8cm,angle=0.]{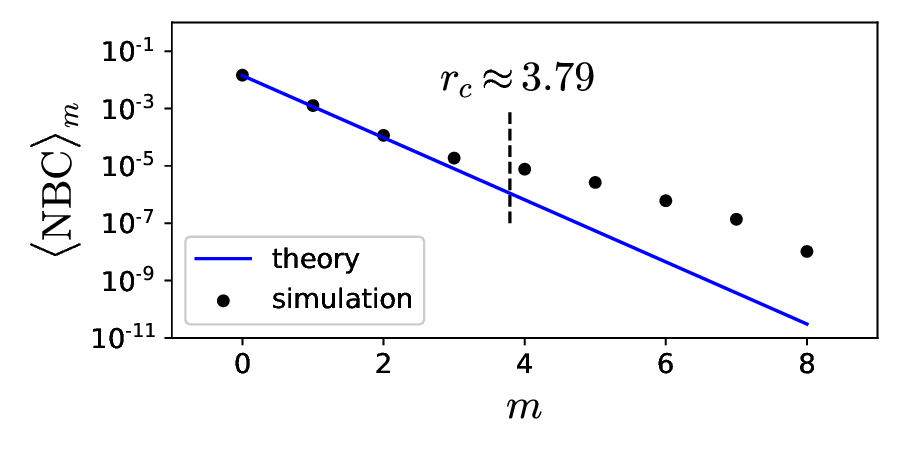}
\caption{Mean NBC as a function of distance from the child network: an ER network of $n=50$ and $z \approx 12.11$ inserted into the Gnutella p2p ($N=62561, b \approx 11.48$) network by merging $n_c = 5$ random nodes in both networks. The range of exponential decay, $r_c$, given by Eq. (\ref{eq42.50}) is marked with a dashed black line. Simulation results were averaged over $100$ realizations.}
\label{fig:decay_ER_in_real}
\end{figure}

\noindent
As an example, Fig. \ref{fig:decay_figures} shows results for the decay of the NBC around the child network in a composite network consisting of the Karate club network inserted, at $n_c=5$ random points, into an ER network of $N=10^6$ and varying mean degree. The LEV of the child network is $\approx 5.29$, therefore in Fig. \ref{fig:decay_figures}(a) localization is absent due to the higher LEV of the mother network. In the cases where localization is present (Fig. \ref{fig:decay_figures}(b,c,d)) the theoretical prediction, Eq. (\ref{eq41.40}) provides a very good fit up to a cutoff distance which depends on the LEV (i.e., the branching, which in this case is closely approximated by the mean degree) of the mother network. As an example of a real-world mother network Fig. \ref{fig:decay_ER_in_real} shows results for a small ER network inserted into the Gnutella p2p network. For small distances away from the child network, the theory works remarkably well.

\noindent
Both in Figs. \ref{fig:decay_figures} and \ref{fig:decay_ER_in_real} the mean NBC noticably deviates, above a certain distance, from the exponential decay predicted by Eq. (\ref{eq41.40}). This is a finite size effect related to the normalization of the NBCs, and can be understood as follows.
Due to the normalization of the NBCs we must have

\begin{align}
1 \approx \langle \rho \rangle \left[ 1 + \frac{f \langle k \rangle}{b} \sum_{m=1}^{\infty} \left( \frac{b}{z} \right)^m \right],
\label{eq41.50}
\end{align}

\noindent
which results in

\begin{align}
\langle \rho \rangle \approx \frac{ z - b }{f \langle k \rangle + z - b},
\label{eq41.60}
\end{align}

\noindent
in accordance with Eq. (\ref{eq32.50}). Equation (\ref{eq41.40}) is approximately valid in an infinite locally tree-like mother network for any $m$ value. The rate of decay of the exponential in Eq. (\ref{eq41.40}) is such that when summing the NBC values of all distances, $m \to \infty$, the normalization condition, Eq. (\ref{eq41.50}), holds. For a finite network, assuming the same $\rho$ value and the same exponential decay rate but a finite maximum distance $m_{\textrm{max}}$, the normalization condition is necessarily violated,

\begin{align}
1 > \langle \rho \rangle \left[ 1 + \frac{f \langle k \rangle}{b} \sum_{m=1}^{ m_{\textrm{max}} } \left( \frac{b}{z} \right)^m \right].
\label{eq41.70}
\end{align}

\noindent
Therefore, to restore correct normalization, for finite mother networks there must exist a cutoff distance above which the NBC decays more slowly. Note that the finite network considered here is still large enough so that $\rho$ is given by the theory derived for infinite mother networks, i.e., we assume that the criterion in Eq. (\ref{eq33.20}) is satisfied.

\subsection{Decay of nonbacktracking centrality in finite graphs}
\label{sec42}

Here we give an approximation to the cutoff distance above which the NBC must decay more slowly than the exponential predicted by Eq. (\ref{eq41.40}).
For simplicity let us assume that the mean NBC decays exponentially up to a cutoff distance $r_c$, and is uniform above that value up to the effective radius

\begin{align}
R &\equiv \frac{ \ln N + \ln(b-1) - \ln \langle k \rangle - \ln n - \ln f }{ \ln b } \nonumber \\
&\approx \frac{ \ln N }{ \ln b }. \label{eq42.05}
\end{align}

\noindent
(Here the effective radius $R$ is the average maximum distance from the child network in a random mother network of size $N$, with mean branching $b$.)
First let us write the mean number of nodes up to distance $r$ in a random mother network,

\begin{align}
\sum_{m=0}^r \langle N_m \rangle \approx \frac{f n \langle k \rangle b^r}{b-1},
\label{eq42.10}
\end{align}

\noindent
and the mean sum of NBCs up to distance $r$, assuming the exponential decay of Eq. (\ref{eq41.20}),

\begin{align}
\sum_{m=0}^r \langle S_m \rangle \approx \langle \rho \rangle + \frac{ f \langle \rho \rangle \langle k \rangle \left[ \left( \frac{b}{z} \right)^r - 1 \right] }{ z \left( \frac{b}{z} - 1 \right) }.
\label{eq42.20}
\end{align}

\noindent
We can write an equation for normalization:

\begin{align}
1 \approx \,\, & \langle \rho \rangle + \frac{ f \langle \rho \rangle \langle k \rangle \left[ \left( \frac{b}{z} \right)^{r_c} - 1 \right] }{ z \left( \frac{b}{z} - 1 \right) } \nonumber \\
&+ \frac{\langle \rho \rangle}{n} z^{-r_c} \left( N - f n \langle k \rangle \frac{b^{r_c}}{b-1} \right),
\label{eq42.30}
\end{align}

\noindent
where the first two terms on the right-hand side give the sum of NBCs up to the cutoff distance $r_c$ from the child network, see Eq. (\ref{eq42.20}). The third term on the right-hand side gives the sum of NBCs of all remaining nodes, that is, nodes at distances larger than $r_c$, considering that NBC from distance $r_c$ onwards is assumed to be uniform, and using Eq. (\ref{eq42.10}).
Accounting for Eq. (\ref{eq32.50}), we have

\begin{align}
\frac{f \langle k \rangle + z - b }{ z-b} \approx \,\, &1 + \frac{ f \langle k \rangle \left[ \left( \frac{b}{z} \right)^{r_c} - 1 \right] }{ z \left( \frac{b}{z} - 1 \right) } \nonumber \\
&+ z^{-r_c} \left( \frac{N}{n} - f\langle k \rangle \frac{b^{r_c}}{b-1} \right).
\label{eq42.40}
\end{align}

\noindent
This results in the cutoff distance

\begin{align}
r_c \approx \frac{ \ln N + \ln (z-b)}{ \ln b } + C,
\label{eq42.50}
\end{align}

\noindent
with

\begin{align}
C = \frac{ \ln (b-1) - \ln (z-1) - \ln \langle k \rangle - \ln (fn)}{ \ln b }.
\label{eq42.55}
\end{align}

\noindent
Expressed using the effective radius $R$,

\begin{align}
r_c \approx R - \frac{ \ln (z-1) - \ln (z-b) }{ \ln b } < R.
\label{eq42.60}
\end{align}

\begin{figure}[H]
\centering
\includegraphics[width=8cm,angle=0.]{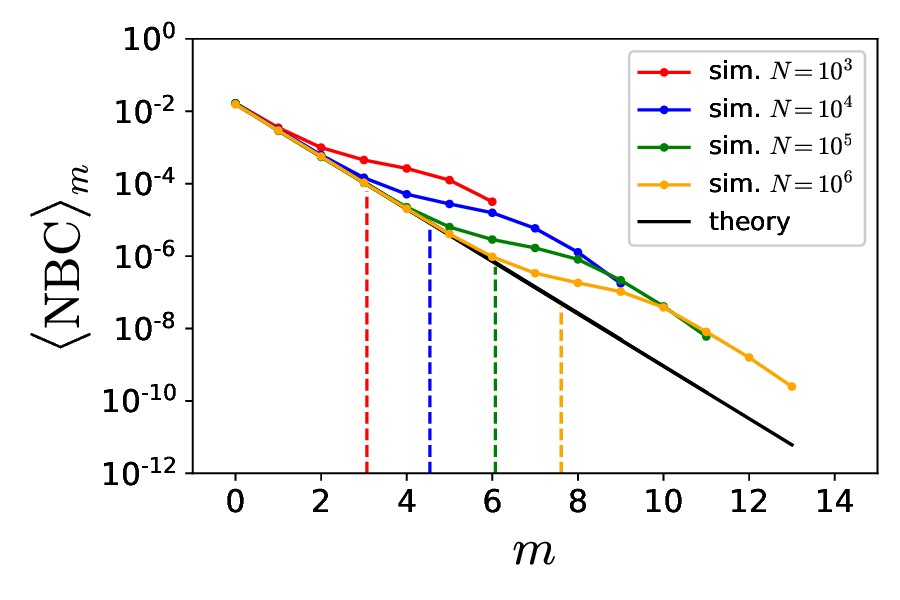}
\caption{The Karate club network ($n=34, z\approx5.29$) inserted into ER networks of mean degree $\langle k \rangle = 4.5 \approx b$ and sizes $N = 10^3, \, 10^4, \, 10^5$ and $10^6$. The insertion was made at $n_c=5$ random points. Solid lines with dots show mean NBC as a function of distance from the child network. The solid black line corresponds to the result, Eq. (\ref{eq41.40}). The range of exponential decay, $r_c$, given by Eq. (\ref{eq42.50}) is marked with dashed lines. Increasing the size of the mother network by a factor of $10$ corresponds to a shift of the range $r_c$ by a value $\ln 10 / \ln 4.5 \approx 1.53$, as predicted by Eq. (\ref{eq42.50}).}
\label{fig:decay_all_N}
\end{figure}

\noindent
We see that $r_c \to R$ as $b \to 1$, i.e., approaching the percolation threshold in the mother network.
Equation (\ref{eq42.50}) slightly overestimates the cutoff distance, as seen in Figs. \ref{fig:decay_figures}, \ref{fig:decay_ER_in_real} and \ref{fig:decay_all_N}. This is due to the fact that the mean NBC displays only a short plateau at the cutoff distance and then continues to decay, contrary to our assumption. The prediction of the cutoff distance, relative to the actual observable value, progressively improves with increasing network size, as shown in Fig. \ref{fig:decay_all_N}, where the Karate club network was inserted into ER mother networks of size ranging from $N = 10^3$ to $10^6$.

\noindent
One can easily show that $r_c$ in Eq. (\ref{eq42.50}) is positive within the range of validity of our theory, Eq. (\ref{eq33.20}), as it must be.

\section{Discussion and conclusions}
\label{sec5}

In this paper we have considered the problem of localization of the NBC in networks where the LEV of the nonbacktracking matrix of a small subgraph is larger than that of the surrounding network. Our results were obtained for a composite network where a completely arbitrary child network is inserted into an infinite locally tree-like mother network. The mother network is allowed to have any correlations, but must be such that the branches emanating from child nodes do not intersect at finite distances. This condition allowed us to utilize the concept of nonbacktracking expansion to derive an exact expression for the sum of NBCs of child nodes. For uncorrelated random mother networks a simple expression was found. For large mother networks the NBCs of child nodes are small (zero in the limit of infinite mother network size) when the LEV of the child network is less than or equal to the LEV of the mother network. Localization on the child network occurs when the child LEV is larger than the mother LEV, and in this case the child node NBCs are strictly positive in the infinite size limit. We show that, in our construction, the LEV of the composite network coincides with the maximum of the two LEVs: that of the child and that of the mother network. This result is a useful addition to what is already known due to the Collatz-Wielandt formula: that the LEV of any network is greater than or equal to the LEV of an arbitrary subgraph of the network. Strict equality applies only in the case of an infinite locally tree-like mother network with non-intersecting branches, however, as we show, the result is accurate also in moderately sized random networks and even in real-world sparse, loopy networks. Our expressions for the NBC of child nodes proved to be good approximations in all these situations.

Similarly to what was seen in the localization of eigenvector centrality (the PEV components of the adjacency matrix), we found an exponential decay of the mean NBC around the child network in the localized state. The rate of the decay, however, is given only by the LEV of the child network, i.e., is independent of the mother network. In infinite systems the exponential decay means that almost all nodes outside the child network have zero NBC. In the language of epidemic spreading in the SIR model, for example, this would mean that disease is only able to spread locally, in the child network or very close to it. In finite networks the exponential decay is only valid up to a certain cutoff distance above which the mean NBC decays more slowly. Our estimate for this cutoff distance agrees well with the simulation results. Interestingly, in simulations of random networks, the mean NBC appears to return to an exponential decay after a short plateau at the predicted cutoff distance. This behaviour, involving peripheral nodes, is not explained by our theory and will require further work to understand.

Our findings contribute to a better understanding of non-recurrent dynamical models, such as the SIR model of epidemic spreading, on certain real-world network structures, where localization may occur due to small, dense subgraphs. Our results were derived for an arbitrary child network, albeit inserted into an infinite locally tree-like mother network. We suggest that the presented ideas will provide a basis for future work on more realistic mother network structures containing finite cycles.

\section*{Acknowledgments}

This work was developed within the scope of the project i3N, UIDB/50025/2020 \& UIDP/50025/2020, financed by national funds through the FCT/MEC--Portuguese Foundation for Science and Technology. G.T. was supported by FCT Grant No. CEECIND/03838/2017.


%

\end{document}